\documentclass[fleqn,10pt]{wlscirep}
\usepackage[utf8]{inputenc}
\usepackage[T1]{fontenc}

\usepackage{url}
\usepackage{graphicx}
\usepackage[table,xcdraw]{xcolor}

\usepackage{lineno}

\title{Urban mobility network centrality predicts social resilience}

\author[1]{Lin Chen}
\author[2,3]{Fengli Xu}
\author[4]{Esteban Moro}
\author[1,5,*]{Pan Hui}
\author[2,3,*]{Yong Li}
\author[6,7,*]{James Evans}
\affil[1]{Department of Computer Science and Engineering, The Hong Kong University of Science and Technology, Hong Kong SAR, P.R. China}
\affil[2]{Department of Electronic Engineering, Tsinghua University, Beijing, P.R. China}
\affil[3]{Beijing National Research Center for Information Science and Technology (BNRist),  Tsinghua University, Beijing, P. R. China}
\affil[4]{Network Science Institute, Department of Physics, Northeastern University, Boston, MA, USA}
\affil[5]{The Hong Kong University of Science and Technology (Guangzhou), Guangzhou, P. R. China}
\affil[6]{Santa Fe Institute, Santa Fe, NM, USA}
\affil[7]{Knowledge Lab and Department of Sociology, University of Chicago, Chicago, IL, USA}
\affil[*]{panhui@ust.hk, liyong07@tsinghua.edu.cn, jevans@uchicago.edu}

\begin{abstract}
Cities thrive on social interactions that foster well-being, innovation, and prosperity; yet, exogenous shocks such as pandemics, hurricanes, and wildfires can severely disrupt them.
Different urban venues exhibit widely divergent response patterns, raising key questions about what factors contribute to these differences and how we can anticipate and respond to them.
Understanding these questions is crucial for safeguarding social resilience, the capacity of urban venues to maintain both visitation and diversity.
In this study, we analyze large-scale human mobility data from 15 US cities covering more than 103 million residents across three distinct urban shocks.
Despite a general trend of declining visitation and weakened social mixing (i.e., intensified segregation) during shocks, our analysis reveals 36.28\%-53.01\% of venues exhibit reduced segregation, and 21.04\%-38.55\% of venues exhibit increased visitation.
By constructing a mobility network interlinking types of urban venues, we reveal that eigenvector network centrality tends to indicate the provision of essential services and robustly predicts social resilience across varied urban shocks.
Specifically, centrality elevates the explanatory power by more than 80\% in predicting both segregation and mobility change, compared with more intuitive features, including pre-shock segregation and mobility.
Furthermore, while core venues (high centrality) and peripheral ones (low centrality) share a similar spatial distribution, they manifest distinct spatiotemporal visitation patterns, with core locations featuring shorter visit distances, broader neighborhood visitation, shorter visitor dwell times, and steadier popularity throughout the day.
Such patterns imply a dual social mechanism, where core venues sustain neighborhood-level social ties through frequent informal interaction, while peripheral ones facilitate deeper engagement around specialized interests and their corresponding social circles.
This suggests that crisis-response efforts should prioritize safeguarding core venue accessibility and their essential services to maintain everyday interaction, while directing targeted support to peripheral venues in order to restore specialized community and exchange following shocks.
By bridging urban mobility research with economic theories that distinguish staple from discretionary products, we propose a \textit{well-and-pool} analogy that suggests how people spend their varying urban mobility budgets. 
In crisis, citizens invest their limited mobility on essentials--like scarce water collecting in a deep well--but when mobility is abundant, like water after rain, people spread across many specialized venues, like water pooling across uneven ground. 
This operational analogy and our findings offer a new, broadly applicable lens on policymaking for urban social resilience and effective crisis response.

\end{abstract}

\begin{document}

\flushbottom
\maketitle
%
%
\thispagestyle{empty}

\section*{Introduction}

Cities are human vibrant hubs~\cite{bettencourt2021introduction} offering abundant opportunities for economic exchange~\cite{bettencourt2007growth}, resource access~\cite{xu2020deconstructing}, and cultural experience~\cite{rahimi2018geography} through a myriad locations, from restaurants and markets to schools and parks. 
These venues serve as `social anchors'~\cite {clopton2011re} that draw together diverse individuals to forge ties, generate innovation~\cite{catalini2018microgeography}, manage conflict~\cite{van2023intergroup}, and improve well-being~\cite{stier2021evidence}.
Recent advances in fine-grained human mobility data make it possible to map these social interactions at an unprecedented scale~\cite{moro2021mobility, nilforoshan2023human,wang2018urban,de2024people,xu2025using}. 
Urban shocks, ranging from pandemics to natural disasters, have become more frequent and disruptive due to climate change and accelerating urbanization, threatening social interaction in ways that weaken community cohesion and slow recovery.

Existing research has documented aggregate patterns of urban mobility in response to shocks, characterized by declining place visitation~\cite{chang2021mobility,li2022spatiotemporal} and heightened socio-demographic segregation~\cite{yabe2023behavioral,napoli2023socioeconomic}.
Nevertheless, these aggregate trends obscure striking variations at the venue level. 
Some places maintain or even increase both visitor volume and demographic diversity, while others collapse in the face of the same disruptive events. 
Yabe et al.~\cite{yabe2025behaviour} study how the impacts of urban shocks propagate across venues, emphasizing their interdependent relationships.
However, we still lack a framework that can anticipate socially resilient venues from \textit{pre-shock} mobility structure, and explain why certain places sustain both visitation and social mixing better than others during shocks.
Filling this gap is essential for crafting effective crisis-response strategies that protect social cohesion, mitigate adversarial impacts~\cite{chen2022strategic}, and accelerate post-shock recovery. 

In this study, we examine how urban venues differ in social resilience to shocks, using large-scale human mobility data from Safegraph across 15 U.S. cities (Metropolitan Statistical Areas, or MSAs) covering more than 103 million residents, 66 types of venues or points of interest (POI) and three kinds of shocks: the COVID-19 pandemic, the 2020 Hurricane ETA, and the 2018 California Wildfire. 
For each venue, we aggregate its visitation records to calculate mobility and segregation change (Figure~\ref{fig1}a, Methods M2).
Our results reveal substantial heterogeneity in venue mobility responses: amidst the general trend of increased segregation and reduced visitation, 36.28\%-53.01\% of venues exhibit reduced segregation, and 21.04\%-38.55\% exhibit increased visitation during shocks (Figure~\ref{fig1}b,c).
Crucially, these changes systematically correlate with venues' sectoral classifications (Figure~\ref{fig1}d). 
Drawing inspiration from economic dependency~\cite{hidalgo2007product}, we construct MSA-scale sector networks to map the preference dependencies between different types of venues (Methods M3).
First, we calculate the visitation preferences of each neighborhood for different venue sectors by comparing neighborhood visitation frequencies to the population average. 
Second, we determine sector-to-sector proximity using conditional probabilities, which capture the likelihood that neighborhoods exhibiting a preference for one venue type also exhibit a preference for another, yielding a population-level coupling between sectors.
We then calculate the eigenvector centrality of each sector within the network. 
Eigenvector centrality captures a recursive notion of importance, assigning higher scores to sectors that are strongly connected to other highly connected types, thereby reflecting their structural prominence in resident mobility pathways.
The resulting sectoral network and centrality measures provide a quantifiable representation of each sector's position within the urban mobility landscape, highlighting interdependencies that emerge from visitor preferences.
With this network, our analysis reveals that sector centrality emerges as a robust predictor of both segregation change and mobility change (Figure~\ref{fig2}), enhancing explanatory power by more than 80\% over traditional metrics such as pre-shock mobility and segregation levels (Figure~\ref{fig3}). 
Furthermore, we identify distinct spatiotemporal patterns of visitation across core and peripheral sectors, where venues from core sectors are marked by shorter visit distances, broader neighborhood visitation, shorter visitor dwell times, and more even popularity throughout the day (Figure~\ref{fig4}).
Such patterns imply a dual social mechanism, where core venues sustain neighborhood-level social ties through frequent informal interactions, while peripheral ones facilitate deeper engagement and support the specialized social circles that congregate there.
These findings suggest that crisis-response efforts should prioritize safeguarding core venue accessibility and essential services to maintain everyday interaction, while directing targeted support to peripheral locations to restore distinctive communities and exchange following shocks.
By bridging urban mobility research with economic theories that distinguish staple from discretionary products~\cite{veblen2017theory,parker2013consumer}, we synthesize our findings into a \textit{well-and-pool} analogy, likening core venues to deep "wells" representing stable, essential services, and peripheral sectors to shallow "pools" that capture volatile, discretionary activities, offering a new lens to interpret, anticipate, and manage urban social resilience.
Overall, this research contributes to a deeper understanding of how collective mobility behavior patterns spontaneously adapt to exogenous shocks, providing actionable insights to enhance the social resilience of urban environments in future crises.

\section*{Results}

\subsection*{Differential Mobility and Segregation Changes Across Venues}
\begin{figure}[ht]
\centering
\includegraphics[width=\linewidth]{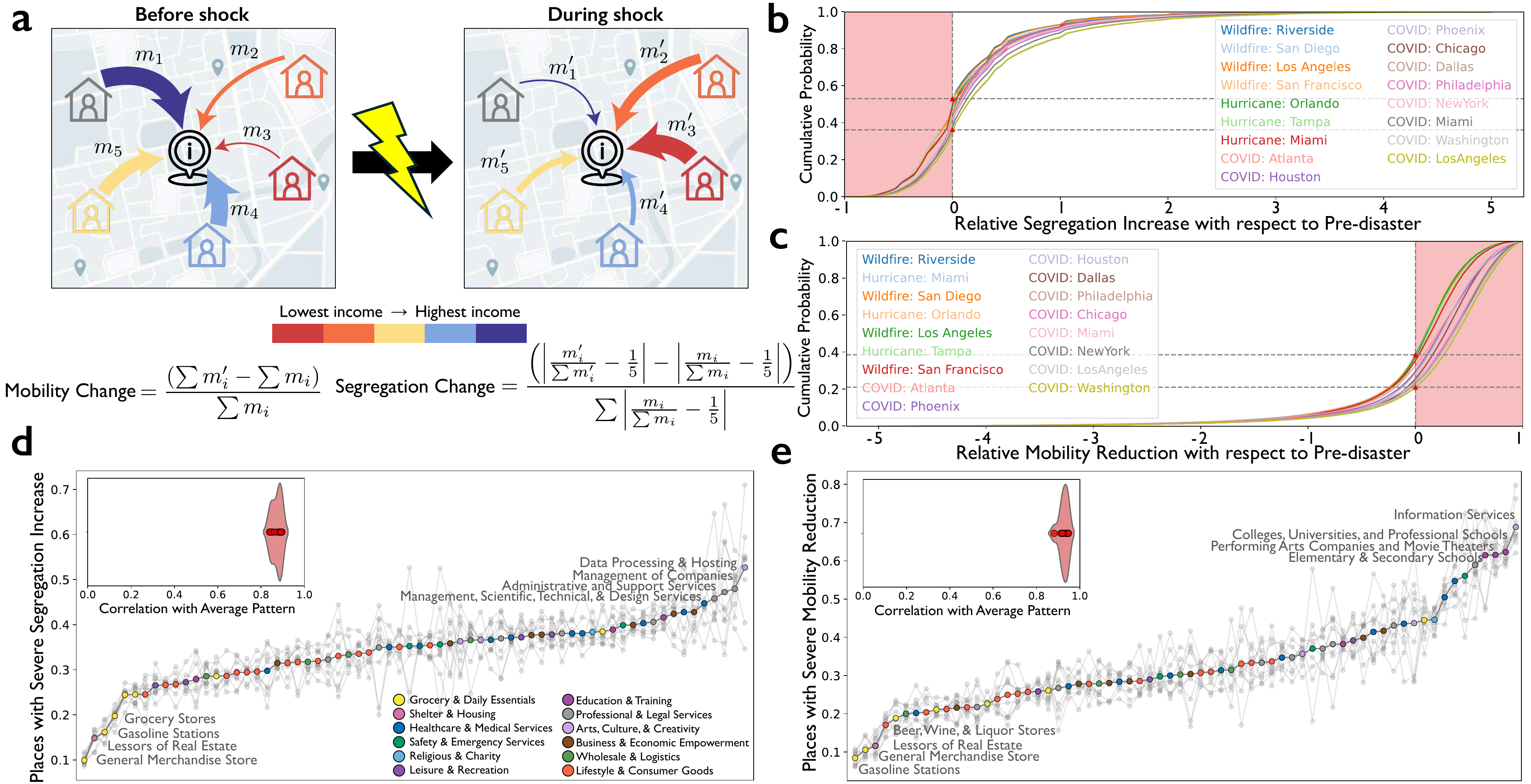}
\caption{\textbf{Social resilience of urban places during shocks.} 
\textbf{a}, Illustrated calculation of mobility and segregation changes during shocks. 
\textbf{b}, Changes in segregation among 15 MSAs during 3 shocks (COVID-19, 2018 California Wildfire, and 2020 Hurricane ETA), with 36.28\%-53.01\% of venues manifesting reduced segregation. 
\textbf{c}, Changes in mobility among 15 MSAs during 3 shocks, with 21.04\%-38.55\% of places showing increased visitation. 
\textbf{d}, Proportions of places in the top 30\% range of segregation increase, aggregated by sector and sorted in ascending order. The middle curves show the mean value across different MSAs. The upper-left subfigure shows the correlation between single-MSA patterns and the mean pattern. The node colors reflect the broader category each sector falls within. 
\textbf{e}, Proportion of places in the top 30\% range of mobility reduction, aggregated by sector and sorted in ascending order. The middle curves show the mean value across different MSAs. The upper-left subfigure shows the correlation between single-MSA patterns and the mean pattern. The node colors reflect the broader category each sector falls within.}
\label{fig1}
\end{figure}

Although previous studies have observed a general trend of increased segregation and reduced visitation among venues in the face of urban shocks~\cite{yabe2023behavioral,chen2023getting}, our analysis uncovers a subset of locations that exhibit notable social resilience.
Across 15 MSAs during three distinct shocks (COVID-19, 2020 Hurricane ETA, and 2018 California Wildfire) in the US, 36.28\%-53.01\% of places are associated with reduced segregation (i.e., increased population mixing) (Figure~\ref{fig1}b), and 21.04\%-38.55\% of places experience increased visitation (Figure~\ref{fig1}c).
Contrasting with the prevailing trend, these venues may play a crucial role in mitigating the negative impacts of shocks, particularly in maintaining social interactions and cohesion. 

To further explore this phenomenon, we then analyzed venue sectoral affiliations derived from the North American Industry Classification System (NAICS) to classify venues according to their industry characteristics or economic activities. 
Specifically, we calculate the proportion of venues in each sector that rank among the top 30\% of those experiencing the greatest segregation deterioration.
Resulting distributions reveal pronounced sectoral differences (Figure~\ref{fig1}d). 
For example, across 10 large MSAs, \textit{General Merchandise Stores} have an average of only 9.9\% of their venues in the severe segregation deterioration region during the COVID-19 pandemic, while \textit{Data Processing Services} exhibit a striking 52.7\% of venues residing in this region, more than five times the former.
Moreover, the Spearman correlation between shock-era segregation patterns observed in individual MSAs and mean pattern across different MSAs is notably strong, with an average of 0.87 ($std=0.02$) and the highest 0.90.
This suggests a high degree of generalizability among sectoral orders across diverse urban contexts.
Similarly, we observe consistent patterns of sectoral mobility change across MSAs, with an average Spearman correlation of 0.69 ($std=0.06$) and a maximum of 0.78 (Figure~\ref{fig1}e).
These findings reinforce the notion that certain types of venues have a greater capacity to attract or maintain populations during shocks, serving as societal anchors and mitigating spatial and social divides exacerbated by adverse events.
As such, they prompt a deeper investigation into the differentiated roles of venues in the urban mobility landscape during urban shocks.


\subsection*{Mobility Network Centrality Predicts Social Resilience}
\begin{figure}[ht]
\centering
\includegraphics[width=\linewidth]{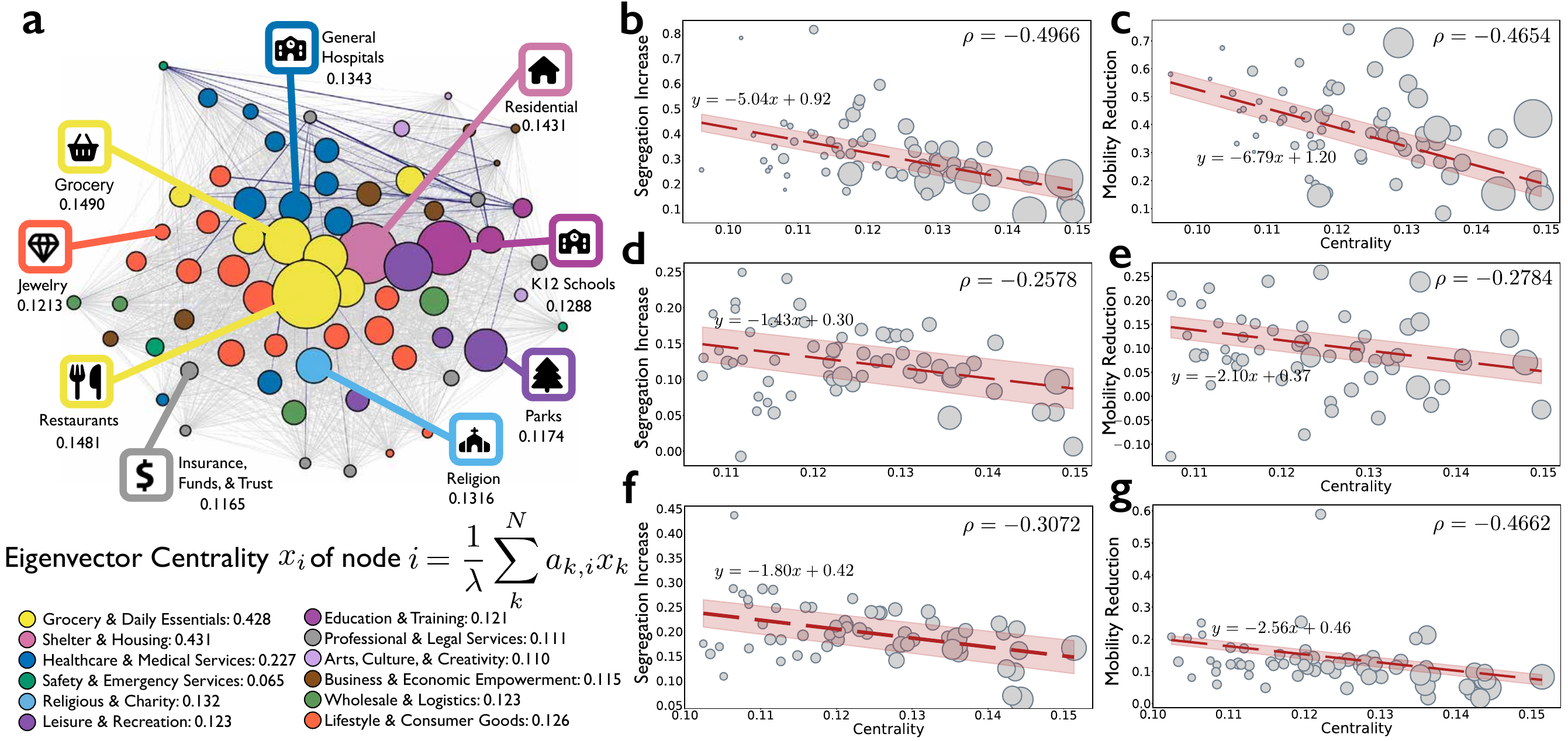}
\caption{\textbf{Uneven segregation and mobility changes across places can be predicted by their centrality in the mobility network.} 
\textbf{a}, Constructed mobility network of places in the pre-disaster period, for the Philadelphia MSA. The node sizes reflect the total visitations to each sector. The node colors reflect the broader category each sector falls within. The value next to each sector/category reflects its average eigenvector centrality.
\textbf{b,d,f}, Relationship between the network centrality of places and their relative segregation change (b) in Philadelphia during COVID-19, (d) in Riverside during the 2018 California Wildfire, and (f) in Miami during the 2020 Hurricane ETA. 
\textbf{c,e,g}, Relationship between the network centrality of places and their relative mobility change (c) in Philadelphia during COVID-19, (e) in Riverside during the 2018 California Wildfire, and (g) in Miami during the 2020 Hurricane ETA. Lines are fitted by OLS regression. Shaded regions indicate 95\% CI.}
\label{fig2}
\end{figure}

\begin{figure}[ht]
\centering
\includegraphics[width=\linewidth]{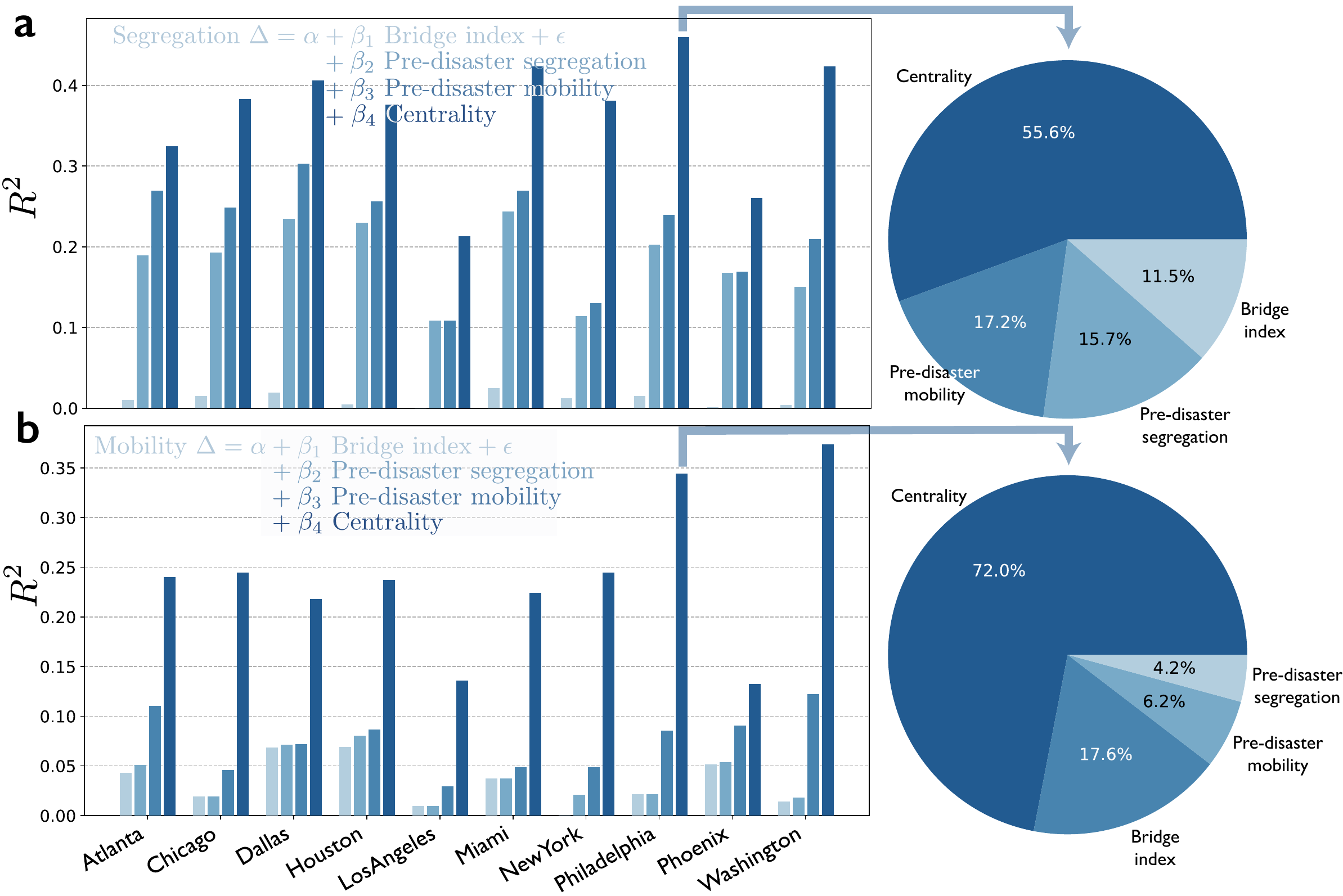}
\caption{\textbf{OLS regression models for segregation and mobility changes during COVID-19.} Pie charts show the relative percentage importance of each feature in Model 4 for the Philadelphia MSA.}
\label{fig3}
\end{figure}

With uneven social resilience across sectors, we aim to quantify this disparity from the perspective of urban visitation networks.
Drawing inspiration from research on economic dependency~\cite{hidalgo2007product}, we first construct a venue sector network for each city based on residents' visitation patterns (see Methods M3).
As shown in Figure~\ref{fig2}a, such a network exhibits a clear hierarchical structure with core-peripheral distinction.
Closer inspection of sector semantics reveals that core nodes in the network correspond to venues that satisfy essential and common needs, such as restaurants and grocery stores, while peripheral nodes correspond to venues that satisfy staple, diverse, or esoteric needs, including information and entertainment services. 

Notably, there is a significant and strong negative correlation between sectors' network centrality and their segregation change (Figure~\ref{fig2}b,d,f), meaning that core places sustain population mixing markedly better than peripheral places.
This relationship holds true across different urban and disaster contexts. Consider the following results from our OLS regression models: during the COVID-19 pandemic in Philadelphia, a unit increase in segregation corresponds to a 5.04-unit decrease in sector centrality; in the Riverside MSA during the 2018 California Wildfire, the corresponding value is 1.43; with 1.80 during the 2020 Hurricane ETA in Miami.
Similarly, we observe a strong and significant negative correlation between the network centrality of sectors and the change in mobility (Figure~\ref{fig2}c,e,g), indicating that core venues sustain visitation attraction during disasters much more effectively than peripheral places. 
Results for other MSAs are presented in the Supplementary Figures.

We provide an interpretation for this phenomenon by drawing a connection to the economic concepts of `staple' versus `discretionary products'~\cite{veblen2017theory,parker2013consumer}: `staple products' refer to essential goods and services necessary for basic needs that typically exhibit stable demand, while `discretionary products' are non-essential goods and services more likely curtailed or postponed during challenging times when financial budgets are tight.
This distinction aligns with our findings on mobility patterns, where, during urban shocks, people's total mobility budgets contract, prompting them to prioritize visits to essential, high-utility venues that satisfy basic needs. 
In this context, we introduce a \textit{well-and-pool} analogy: just as water collects in a deep well during a drought, individuals during a crisis focus their limited mobility resources on core venues—those with high network centrality—sustaining critical neighborhood-level interactions. 
When mobility is abundant, akin to water pooling after rain, individuals are more likely to spread across a wider range of venues, including those serving specialized or discretionary needs. 
This analogy captures how crises lead to shifts in mobility priorities, with core venues acting as the "well" that maintains essential social ties, while peripheral venues represent the "pool" where more discretionary interactions occur when conditions allow.


Although sector centrality demonstrates a strong correlation with both segregation and mobility change, it is possible that this may simply capture information already accounted for by intuitive features, such as pre-disaster segregation or mobility levels.
To examine this possibility, we construct a series of ordinary least squares (OLS) regression models to estimate the explanatory power of distinct variable combinations.
Our baseline covariates include two pre-disaster features—each sector’s pre-disaster segregation level and pre-disaster mobility level—as well as a bridge index, which provides a static, location-based estimate of a sector’s capacity to connect people from different income groups through the spatial catchments of its venues~\cite{nilforoshan2023human}.
We standardize the input features to ensure stable outputs and facilitate effect size comparisons.
Taking Philadelphia as an example, we begin by examining the impact of pre-disaster conditions on segregation change (Table~\ref{tab:regression_segchange}). 
Pre-disaster segregation alone explains little of the sector-level variation in segregation change ($R^2=0.0151$, Column 1). Adding pre-disaster mobility increases the explained variance to $R^2=0.2023$ (Column 2), and incorporating the bridge index yields only a modest additional gain to $R^2=0.2392$ (Column 3). By contrast, introducing sector centrality substantially improves model fit, raising the explained variance to $R^2=0.4599$ (Column 4), nearly doubling the explanatory power relative to the best baseline specification.
In stark contrast, the introduction of sector centrality into the regression model leads to a drastic improvement, doubling the explained variation to 45.99\% (Column 4).
This highlights the substantial contribution of sector centrality in explaining changes in segregation, suggesting that sectors closer to the core of the mobility network sustain better population mixing during disruptions.
Specifically, a unit increase in sector centrality corresponds to a 0.0709-unit decrease in segregation change ($p<0.001$, two-sided Student’s t-test).

Similarly, for mobility change (Table~\ref{tab:regression_mobchange}), baseline specifications relying on pre-disaster segregation, pre-disaster mobility, and the bridge index explain little variance ($R^2<0.09$, Columns 1–3), whereas adding sector centrality increases the explained variance to $R^2=34.44$ (Column 4).
In this case, a unit increase in sector centrality corresponds to a 0.0945-unit decrease in mobility change (p < 0.001), suggesting that sectors closer to the mobility network core are more successful in maintaining their appeal during disruptions.
Together, these findings reveal that sector centrality is not simply a proxy for pre-existing conditions such as segregation or mobility levels, but rather provides unique insights into the dynamics of social resilience. 
In other words, sectors more centrally located in the urban mobility network are generally more socially resilient, both in maintaining diverse social interaction and visitation activity.
Regression results for other MSAs and shocks can be found in Figure~\ref{fig3}, and Supplementary Tables 1-32.
In conclusion, while pre-disaster segregation and mobility levels may explain part of the observed changes in segregation and mobility, sector centrality accounts for more.
Core places exhibit greater social resilience, maintaining better population mixing and more sustained visitor engagement during periods of shock.


\subsection*{Core and Peripheral Venues}
\begin{figure}[ht]
\centering
\includegraphics[width=\linewidth]{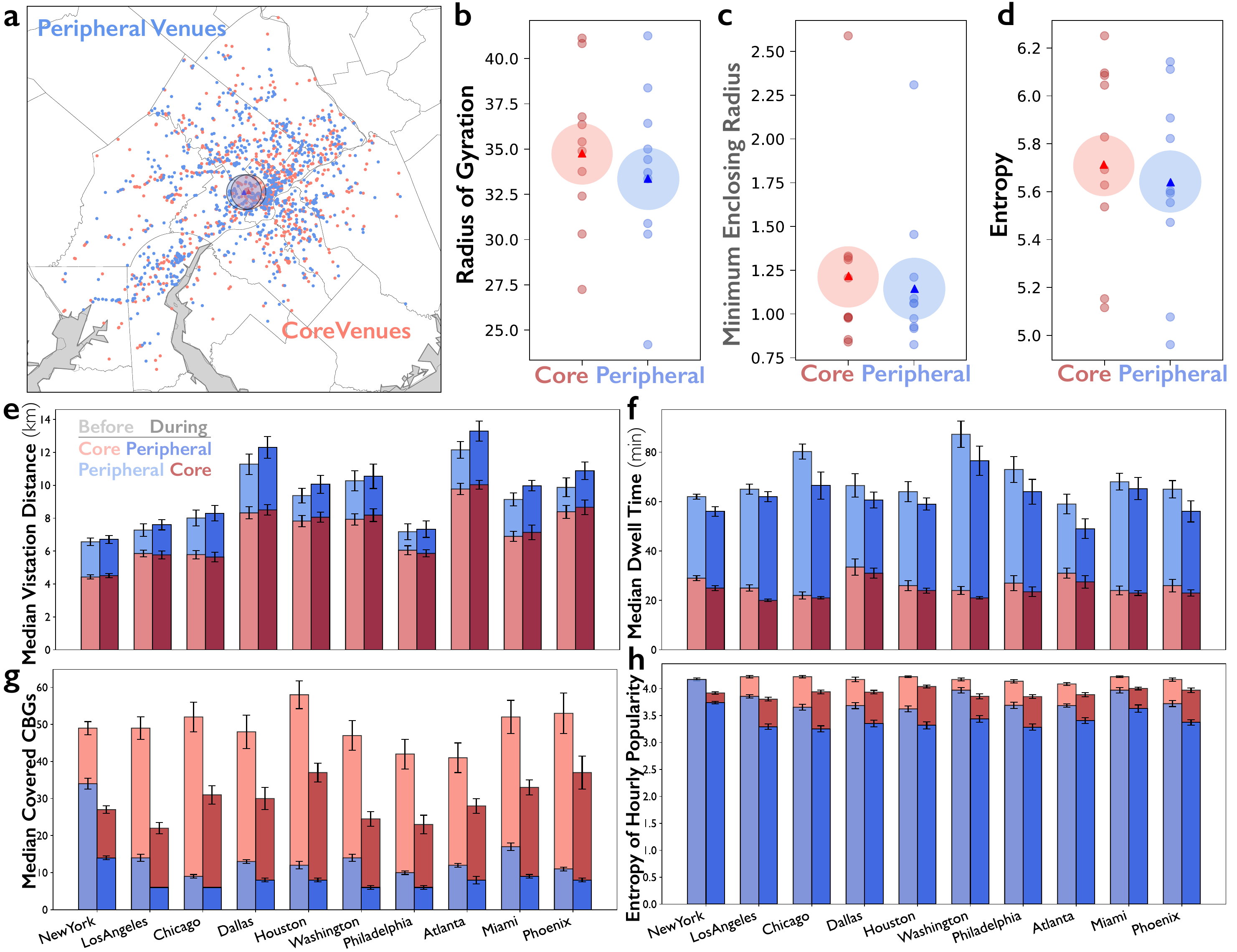}
\caption{\textbf{Comparison between core and peripheral venues.} \textbf{a}, Geographical distribution of core (red dots) and peripheral (blue dots) venues in the Philadelphia MSA. \textbf{b-d}, Comparison of distribution statistics between core and peripheral venues in 10 large MSAs. Each dot represents one MSA. \textbf{e}, Median visitation distances to core and peripheral venues before and during COVID-19. \textbf{f}, Median number of covered CBGs of core and peripheral venues before and during COVID-19. \textbf{g}, Median visitor dwell time at core and peripheral places before and during COVID-19. \textbf{h}, Entropy of hourly popularity of core and peripheral places before and during COVID-19. Error bars indicating 95\% CI.}
\label{fig4}
\end{figure}

To gain a deeper understanding of how mobility network centrality distinguishes venues, we conduct a series of comparison analyses between core and peripheral venues, focusing on both their static geographic distribution and dynamic visitation patterns.
We define core and peripheral sectors as the top 10 and bottom 10 sectors, respectively, ranked by sectoral network centrality.
To ensure balanced comparisons, we included all POIs in peripheral sectors and randomly sampled an equal number of POIs from the larger pool of POIs in core sectors.
First, we visualize the geographical distribution of core and peripheral venues.
As shown in Figure~\ref{fig4}a, which takes Philadelphia as an example, no differences are observable. 
We further use three complementary metrics to quantify their spatial distributions (Figure~\ref{fig4}b-d).
The first, the Radius of Gyration (RoG), measures the average distance venues are distributed from their average location.
The second metric, Minimum Enclosing Radius (MER), measures the smallest radius within which all venue locations are contained, capturing the spatial footprint of places.
The third metric, Entropy, measures the unevenness of venue geographical distributions.
We observe no significant differences between core and peripheral venues along these three metrics (RoG: $p=0.5863$; MER: $p=0.7926$; Entropy: $p=0.5089$; All via two-sided Student’s t-test).
This result suggests that observed differences in network centrality do not stem from spatial configuration.
Instead, centrality differences arise from their function within human mobility pathways, i.e., how and when people visit them.

We now proceed to examine dynamic visitation patterns, focusing on spatial and temporal dimensions that help explain why core venues exhibit more robust social resilience in the face of shocks.
Statistical comparisons across MSAs use a two-sided Wilcoxon signed-rank test on within-MSA medians, and we report the corresponding effect size ($r$).
In terms of spatial reach, we observe that the median travel distances to core places ($median=7.36km$) are consistently shorter than those to peripheral places ($median=9.26km$) across MSAs ($p=0.002$, $r=0.886$; Figure~\ref{fig4}e).
Aligning with research on urban walking behaviors~\cite{yang2012walking}, shorter visitation distances to core venues could reflect their greater connectivity and relevance to a larger segment of the population, often fulfilling everyday needs within reachable distances. 
In contrast, peripheral places require longer travel distances, suggesting that they are more specialized and cater to niche needs, attracting fewer but more specific visitors who are willing to travel farther for the services offered.
During shocks, median travel distances showed a small increase for core venues ($median = +2.41\%$, $p = 0.131$, $r = 0.50$) and a larger increase for peripheral venues ($median = +5.80\%$, $p = 0.002$, $r = 0.886$).
Nevertheless, the travel distance to core venues remains shorter than that to peripheral ones ($p=0.002$, $r=0.886$).
Moreover, the median number of neighborhoods served by a core venue ($median=49$) is significantly higher than that by a peripheral location ($median=13$), consistent across MSAs (Figure~\ref{fig4}f).
Serving more neighborhoods effectively enables core venues to function as shared community hubs, ensuring they still attract a critical mass of visitors even as overall mobility declines.
During shocks, both core and peripheral venues cover significantly fewer neighborhoods, but core venues continue to serve more than peripheral ones.

For temporal variation, we observe that the median dwell time at core venues ($median=26.00min$) is notably shorter than that at peripheral ones ($median=65.75min$) across MSAs ($p=0.002$, $r=0.886$; Figure~\ref{fig4}g).
This is consistent with intuition, where visits to core venues---such as grocery stores---are typically task-oriented and efficiency-driven, aiming to fulfill routine, essential needs in a relatively short order.
In contrast, visits to peripheral venues tend to be less time-sensitive, such as entertainment, leisure, or relating to focused personal interests and leading to naturally longer stays.
Shorter visit durations are often perceived as safer, as they reduce potential health risks associated with exposure, and they fit more readily into constrained schedules when mobility is restricted. 
Thus, a lower commitment threshold encourages people to undertake these quick errands, and the resulting high visitor turnover enables continued micro-encounters across diverse groups during times of shock.
Furthermore, we find that the entropy of hourly visitor popularity for core venues is slightly higher ($median = 4.17$) compared to peripheral places ($median = 3.70$) across MSAs ($p=0.004$, $r=0.889$; Figure~\ref{fig4}h). 
Higher entropy for core venues indicates that the demand for these services is spread over a wider range of time, possibly because they serve as frequent touchpoints in residents' daily routines.
In turn, stable and round-the-clock demand makes these places more robust to disruptions~\cite{d2018role}.
These differences are further accentuated during shocks, reflecting how demand for essential services remains more stable throughout the day, whereas the pursuit of peripheral services is more sporadic and concentrates at certain times.
Analysis results for other shocks are presented in the Supplementary Figures, confirming these trends across various scenarios and demonstrating the robustness of our findings.

Taken together, these findings emphasize that observed differences in network centrality largely stem from how and when core versus peripheral venues are utilized, and not their location. 
The combination of shorter travel distances, wider neighborhood visitation, higher visitor turnover, and more stable daily demand underpins the social resilience of core venues by keeping them functionally essential and socially significant.

\section*{Discussion}

While urban shocks generally lead to increased segregation and reduced visitation, a substantial proportion of venues maintain or even increase the likelihood and frequency of social mixing. 
Our study reveals a critical factor underlying the heterogeneity in how urban environments and their venues respond to shocks.
By constructing a preference dependency network between sectors, we find that places with high network centrality are significantly more likely to sustain visitation and social integration during urban shocks.
This finding highlights the existence of socially resilient places that continue to foster social cohesion amidst disruption.
Furthermore, our network-based perspective shows that centrality can predict social resilience significantly more accurately than prior metrics.
From an urban planning standpoint, acknowledging this new metric enables more targeted and effective intervention.
City planners could focus on strengthening core venues by ensuring robust and inclusive transportation infrastructure~\cite{xu2023interconnectedness}, so that venue locations remain operational even when shock events restrict mobility.
Meanwhile, safeguarding peripheral venues through actions like targeted economic relief~\cite{bartik2020impact} and flexible zoning for mixed-use development~\cite{salat2017systemic} can help retain cultural and economic diversity.

The distinct spatiotemporal visitation patterns of core and peripheral venues provide different flavors of social interactions for urban residents.
Core venues, such as grocery stores and restaurants, support frequent, short-duration encounters that reinforce neighborhood-level social ties.
These informal yet regular interactions mirror what Jane Jacobs describes as the "public ballet" of city life, where routine social exchange contributes to urban vibrancy and collective well-being~\cite{jacobs1961death}.
Although each encounter may be brief, the cumulative effect of these interactions weaves many ``weak ties''~\cite{granovetter1973strength} that keep social networks active.
Prior research suggests that such spontaneous encounters enhance collective intelligence by facilitating diverse and serendipitous idea-sharing~\cite{woolley2010evidence}.
In contrast, peripheral venues, such as entertainment locations and specialty services, tend to involve longer-distance travel and purpose-driven visits. 
Prior research has shown that long-distance visits are often motivated by pre-existing social ties, such as friendships, rather than chance encounters with strangers~\cite{cho2011friendship}.
Being less spontaneous, these interactions provide important opportunities for in-depth engagement around focused activities, such as in-person conversations among researchers to catalyze scientific breakthroughs and technological innovation~\cite{lin2023remote}.
Together, these findings highlight a dual process underlying urban social dynamics: core venues sustain a broad base of social connectivity through frequent, low-intensity interactions, while peripheral venues facilitate deeper, high-intensity engagement within specific social circles.
During urban shocks, this balance shifts toward low-intensity interactions at core venues, as individuals prioritize essential needs and reduce discretionary mobility.
This temporary realignment may have positive secondary effects on urban adaptation.
For example, increased reliance on local core venues could foster stronger neighborhood networks, reinforcing mutual aid systems and localized social support.
Research on disaster recovery suggests that informal local ties play a crucial role in collective resilience, enabling faster responses and more efficient resource sharing~\cite {aldrich2012building}.
Rather than a mere restriction, spontaneous shifts toward low-intensity interactions during urban shocks can be understood as a collective adaption that preserves social connectivity, strengthens local resilience, and fosters long-term urban evolution.

The \textit{well-and-pool} analogy we propose---linking core and peripheral venues to essential versus discretionary services---offers a novel conceptual framework for understanding the spontaneous dynamics of urban resilience. During periods of crisis, citizens invest their limited mobility on essentials--like scarce water draining to the bottom of a deep well. By contrast, when mobility is abundant, like water after rain, people spread across many specialized venues, well beyond the well, like water collecting in a myriad shallow pools.
This analogy highlights a crucial policy insight: preserving social integration during urban shocks requires ensuring the resilience of core venues, as their stability enables cities to maintain fundamental mobility, social interaction, and access to resources.
By drawing on economic theories that distinguish staple from discretionary products, we explain the resilience of core venues in terms of their essential nature. 
This analogy provides a theoretical framework to understand the patterns of mobility we observe and emphasizes the stability of essential services during crises.
At the same time, the \textit{well-and-pool} analogy highlights the dynamic interdependence between core and peripheral venues: while core locations sustain fundamental interaction, peripheral ones contribute to long-term urban vitality and diversification, making them critical for post-shock recovery and future urban adaptation.

When interpreting our results, several limitations should be considered.
While our analysis spans multiple cities and shock events, it is still possible that the patterns we observe may not fully generalize to all urban contexts or types of shocks.
Future research could expand the analysis to a broader range of cities, including those outside the US and those with different socio-demographic dynamics.
Moreover, although we identify sectoral centrality as a robust indicator of social resilience, other factors might also influence it, such as local economic conditions and the accessibility of transportation infrastructure.
Future studies could explore these factors in greater detail, potentially integrating diverse data sources to enhance the understanding of mechanisms behind urban resilience.
Our income stratification relies on CBG-level median household income, which may mask within-CBG heterogeneity and thereby attenuate income-related gradients in segregation. 
In addition, phone-based mobility data may differentially capture populations across demographic groups, and misclassification or miss listings may introduce measurement noise in sector-level summaries~\cite{grigoropoulou2025large}.
Finally, our research focuses on the immediate effects of urban shocks on mobility behavior and social mixing patterns, which may not fully capture long-term recovery dynamics.
Future research could investigate whether transient disturbances differ from lasting effects that may alter social mixing patterns over the long run. Nevertheless, this investigation reveals the powerful ways in which mobility patterns in peace and prosperity provide the key to how mobility and visitation change during seasons of disruption.


\section*{Methods}

\subsection*{M1. Mobility and Demographic Data} \label{method:data}
To analyze urban mobility behavior, we utilize Safegraph Patterns, a large-scale human mobility dataset collected from millions of US mobile phone users with consent\footnote{https://www.safegraph.com/}.
The dataset provides aggregate information on monthly visits to venues (i.e., points of interest, or POIs).
To ensure user privacy, SafeGraph employs aggregation and anonymization techniques that prevent the identification of individuals\footnote{https://www.safegraph.com/privacy-policy}. 
Specifically, all visit data is aggregated at the census block group (CBG) level, preventing the tracing of individual users.
Furthermore, SafeGraph applies privacy thresholds to exclude data points from locations with insufficient visitor counts, thereby reducing the risk of re-identification while enabling investigation into human mobility patterns.
We also utilize the Safegraph Places dataset\footnote{https://www.safegraph.com/products/places} to obtain the location (longitude and latitude) and NAICS category information of POIs.
We map POIs to NAICS categories and construct 66 venue types via a curated grouping scheme. 
When a 4-digit NAICS category had insufficient coverage for stable estimation, we aggregated it to its parent 3-digit category; all subsequent analyses use these harmonized sector definitions.

To understand visitor distributions, we utilize the 2019 American Community Survey (ACS) 5-year Estimates.
This dataset provides population characteristics at the CBG level, which is the smallest geographic unit with publicly available demographic statistics and typically encompasses between 600 and 3,000 residents. 
For each MSA, we divide CBGs into five income groups, each representing approximately equal portions of the population, based on their median household income, and calculate segregation at POIs according to these groups.

\subsection*{M2. Quantification of Segregation and Mobility Change} \label{method:change_quantification}
We quantify experienced income segregation as the deviation of observed visitation patterns from an idealized scenario where visitors from different income groups visit a venue (POI) with equal frequency, following previous research~\cite{moro2021mobility}.
Specifically, segregation $S$ at a POI is computed as: 
\begin{equation}
    S = \tau |\frac{m_i}{\sum m_i} - \frac{1}{n}|,
\end{equation}
where $m_i$ represents visitations from the $i$-th income group, $n$ represents the number of income groups, and $\tau$ is a normalization constant ensuring that $S$ remains within the range $[0,1]$.
A lower $S$ value indicates greater population mixing, while a higher $S$ value suggests strong segregation.
To assess behavioral changes at each venue during a shock, we calculate both visitation levels and income segregation during the period of shock, and compare them against the corresponding values from an equally long pre-shock period (immediately preceding the shock). 
The relative changes in segregation ($\Delta S$) and visitation ($\Delta M$) at a POI are thus computed as:
\begin{equation}
    \Delta S = \frac{S_{in\_shock} - S_{pre\_shock}}{S_{pre\_shock}},
\end{equation}
\begin{equation}
    \Delta M = \frac{M_{in\_shock} - M_{pre\_shock}}{M_{pre\_shock}}.
\end{equation}

\subsection*{M3. Construction of Sector Network} \label{method:network_construction}
Inspired by the concept of Revealed Comparative Advantage (RCA) in economic dependency~\cite{hidalgo2007product}, we construct a venue or POI sectoral network for each metropolitan statistical area (MSA) based on the visitation by residents.
In this network, each node represents a venue sector, i.e., the type of POIs providing similar functions, and each edge represents the co-visitation revealed preferences of two POI sectors.
Constructing the network requires determining the weight of edges between any two POI sectors.
This process involves two steps:

\textbf{Step 1: Calculating CBG-Sector Preferences.} 
We begin by computing how much each CBG prefers a given sector compared to the general population.
Let $f_{m, i}$ denote the visitation frequency of CBG $m$ to sector $i$, and $f_{i}$ denote the average visitation frequency of the whole population to sector $i$.
We calculate the visitation preferences $R_{m,i}$ of each CBG $m$ in visiting each sector $i$ as:

\begin{equation}
    R_{m, i} = \frac{f_{m, i}}{f_{i}}.
\end{equation}

We consider a sector $i$ preferred by a CBG if $R_{m, i} > 1$, indicating that residents of that CBG visit the sector more frequently than the general population.

\textbf{Step 2: Calculating Sector-Sector Proximity.} 
Next, we calculate the proximity $P_{i, j}$ between sector $i$ and sector $j$, which measures the likelihood that if one sector is preferred, the other is also preferred. 
We define: 

\begin{equation}
    P_{i, j} = \text{max}\{P(R_i|R_j), P(R_j|R_i)\},
\end{equation}

where $P(R_i|R_j)$ represents the conditional probability that sector $i$ is preferred, given that sector $j$ is also preferred, evaluated across all CBGs.
The final proximity value is taken as the maximum of the two conditional probabilities, ensuring that we capture the stronger relationship linking each pair of venue sectors.



\section*{Data Availability}

Safegraph Patterns data can now be obtained from Advan through a paid subscription (\url{https://www.deweydata.io/data-partners/advan}).
Safegraph Places data can be obtained from Safegraph (\url{https://www.safegraph.com/products/places}).
CBG demographic data is obtained from the official website of American Community Surveys (\url{https://www.census.gov/programs-surveys/acs/}).

\section*{Code Availability}
The Python code to reproduce the main results in this paper will be made public in \url{https://github.com/LinChen-65/social-resilience}.

\bibliography{sample}


\section*{Author contributions}

F.X., Y.L., and J.E. conceived the project and provided the research outline. 
L.C., F.X., and Y.L. designed the research methods.
L.C. conducted the experiments.
L.C. and J.E. prepared the figures.
P.H., E.M., and J.E. provided critical revisions.
All authors analyzed the results and participated in the writing of the manuscript.

\section*{Competing Interests}
J.E. has a commercial affiliation with Google, but Google had no role in the design, analysis, or decision to publish this study. The other authors declare no other competing interests.

\section*{Additional information}
Supplementary Information is available for this manuscript.

\clearpage

\begin{table}[ht]
\centering
\caption{Regression table for segregation change, Philadelphia, COVID-19.}
\label{tab:regression_segchange}
\resizebox{0.6\textwidth}{!}{%
\begin{tabular}{|lcccc|}
\hline
\rowcolor[HTML]{225B91} 
\multicolumn{5}{|c|}{\cellcolor[HTML]{225B91}{\color[HTML]{FFFFFF} Dependent variable: segregation change during disaster}}                                                                                                                                                         \\ \hline
\rowcolor[HTML]{CCD2D8} 
\multicolumn{1}{|l|}{\cellcolor[HTML]{CCD2D8}}                         & \multicolumn{1}{c|}{\cellcolor[HTML]{CCD2D8}(1)}     & \multicolumn{1}{c|}{\cellcolor[HTML]{CCD2D8}(2)} & \multicolumn{1}{c|}{\cellcolor[HTML]{CCD2D8}(3)} & (4)                                           \\ \hline
\rowcolor[HTML]{E7EAED} 
\multicolumn{1}{|l|}{\cellcolor[HTML]{E7EAED}Pre-disaster segregation} & \multicolumn{1}{c|}{\cellcolor[HTML]{E7EAED}-0.0145} & \multicolumn{1}{c|}{\cellcolor[HTML]{E7EAED}-0.0283$^{**}$}    & \multicolumn{1}{c|}{\cellcolor[HTML]{E7EAED}-0.0258$^{*}$}    &  \multicolumn{1}{c|}{\cellcolor[HTML]{E7EAED}-0.0434$^{***}$}                                             \\ \hline
\rowcolor[HTML]{CCD2D8} 
\multicolumn{1}{|l|}{\cellcolor[HTML]{CCD2D8}}                         & \multicolumn{1}{c|}{\cellcolor[HTML]{CCD2D8}p=0.325} & \multicolumn{1}{c|}{\cellcolor[HTML]{CCD2D8}p=0.043}    & \multicolumn{1}{c|}{\cellcolor[HTML]{CCD2D8}p=0.062}    &  \multicolumn{1}{c|}{\cellcolor[HTML]{CCD2D8}p=0.001}                                             \\ \hline
\rowcolor[HTML]{E7EAED} 
\multicolumn{1}{|l|}{\cellcolor[HTML]{E7EAED}Pre-disaster mobility}    & \multicolumn{1}{c|}{\cellcolor[HTML]{E7EAED}-}       & \multicolumn{1}{c|}{\cellcolor[HTML]{E7EAED}-0.0528$^{***}$}    & \multicolumn{1}{c|}{\cellcolor[HTML]{E7EAED}-0.0485$^{***}$}    & \multicolumn{1}{c|}{\cellcolor[HTML]{E7EAED}-0.0122}                                              \\ \hline
\rowcolor[HTML]{CCD2D8} 
\multicolumn{1}{|l|}{\cellcolor[HTML]{CCD2D8}}                         & \multicolumn{1}{c|}{\cellcolor[HTML]{CCD2D8}-}       & \multicolumn{1}{c|}{\cellcolor[HTML]{CCD2D8}p<0.001}    & \multicolumn{1}{c|}{\cellcolor[HTML]{CCD2D8}p=0.001}    &  \multicolumn{1}{c|}{\cellcolor[HTML]{CCD2D8}p=0.380}                                             \\ \hline
\rowcolor[HTML]{E7EAED} 
\multicolumn{1}{|l|}{\cellcolor[HTML]{E7EAED}Bridge index}             & \multicolumn{1}{c|}{\cellcolor[HTML]{E7EAED}-}       & \multicolumn{1}{c|}{\cellcolor[HTML]{E7EAED}-}    & \multicolumn{1}{c|}{\cellcolor[HTML]{E7EAED}-0.0230$^{*}$}    &  \multicolumn{1}{c|}{\cellcolor[HTML]{E7EAED}-0.0226$^{**}$}                                             \\ \hline
\rowcolor[HTML]{CCD2D8} 
\multicolumn{1}{|l|}{\cellcolor[HTML]{CCD2D8}}                         & \multicolumn{1}{c|}{\cellcolor[HTML]{CCD2D8}-}       & \multicolumn{1}{c|}{\cellcolor[HTML]{CCD2D8}-}    & \multicolumn{1}{c|}{\cellcolor[HTML]{CCD2D8}p=0.088}    &  \multicolumn{1}{c|}{\cellcolor[HTML]{CCD2D8}p=0.049}                                              \\ \hline
\rowcolor[HTML]{E7EAED} 
\multicolumn{1}{|l|}{\cellcolor[HTML]{E7EAED}Centrality}               & \multicolumn{1}{c|}{\cellcolor[HTML]{E7EAED}-}       & \multicolumn{1}{c|}{\cellcolor[HTML]{E7EAED}-}    & \multicolumn{1}{c|}{\cellcolor[HTML]{E7EAED}-}    &  \multicolumn{1}{c|}{\cellcolor[HTML]{E7EAED}\textbf{-0.0709$^{***}$}}                                             \\ \hline
\rowcolor[HTML]{CCD2D8} 
\multicolumn{1}{|l|}{\cellcolor[HTML]{CCD2D8}}                         & \multicolumn{1}{c|}{\cellcolor[HTML]{CCD2D8}-}       & \multicolumn{1}{c|}{\cellcolor[HTML]{CCD2D8}-}    & \multicolumn{1}{c|}{\cellcolor[HTML]{CCD2D8}-}    & \multicolumn{1}{c|}{\cellcolor[HTML]{CCD2D8}\textbf{p<0.001}}                                              \\ \hline
\rowcolor[HTML]{E7EAED} 
\multicolumn{1}{|l|}{\cellcolor[HTML]{E7EAED}Const}                    & \multicolumn{1}{c|}{\cellcolor[HTML]{E7EAED}0.2950$^{***}$}  & \multicolumn{1}{c|}{\cellcolor[HTML]{E7EAED}0.2950$^{***}$}    & \multicolumn{1}{c|}{\cellcolor[HTML]{E7EAED}0.2950$^{***}$}    & \multicolumn{1}{c|}{\cellcolor[HTML]{E7EAED}0.2950$^{***}$}                                              \\ \hline
\rowcolor[HTML]{CCD2D8} 
\multicolumn{1}{|l|}{\cellcolor[HTML]{CCD2D8}}                         & \multicolumn{1}{c|}{\cellcolor[HTML]{CCD2D8}p<0.001} & \multicolumn{1}{c|}{\cellcolor[HTML]{CCD2D8}p<0.001}    & \multicolumn{1}{c|}{\cellcolor[HTML]{CCD2D8}p<0.001}    & \multicolumn{1}{c|}{\cellcolor[HTML]{CCD2D8}p<0.001}                                              \\ \hline
\rowcolor[HTML]{E7EAED} 
\multicolumn{1}{|l|}{\cellcolor[HTML]{E7EAED}R$^2$}                       & \multicolumn{1}{l|}{\cellcolor[HTML]{E7EAED}0.0151}  & \multicolumn{1}{l|}{\cellcolor[HTML]{E7EAED}0.2023}    & \multicolumn{1}{l|}{\cellcolor[HTML]{E7EAED}0.2392}    & \multicolumn{1}{l|}{\cellcolor[HTML]{E7EAED}\textbf{0.4599}} \\ \hline
\rowcolor[HTML]{CCD2D8} 
\multicolumn{1}{|l|}{\cellcolor[HTML]{CCD2D8}Adjusted R$^2$}              & \multicolumn{1}{l|}{\cellcolor[HTML]{CCD2D8}-0.0003} & \multicolumn{1}{l|}{\cellcolor[HTML]{CCD2D8}0.1770}    & \multicolumn{1}{l|}{\cellcolor[HTML]{CCD2D8}0.2024}    & \multicolumn{1}{l|}{\cellcolor[HTML]{CCD2D8}\textbf{0.4245}} \\ \hline
\multicolumn{5}{c}{\footnotesize Note: $^{***}$ p<0.01, $^{**}$ p<0.05, $^{*}$ p<0.1.}\\
\end{tabular}%
}
\end{table}

\begin{table}[ht]
\centering
\caption{Regression table for mobility change, Philadelphia, COVID-19.}
\label{tab:regression_mobchange}
\resizebox{0.6\textwidth}{!}{%
\begin{tabular}{|lcccc|}
\hline
\rowcolor[HTML]{225B91} 
\multicolumn{5}{|c|}{\cellcolor[HTML]{225B91}{\color[HTML]{FFFFFF} Dependent variable: mobility change during disaster}}                                                                                                                                                         \\ \hline
\rowcolor[HTML]{CCD2D8} 
\multicolumn{1}{|l|}{\cellcolor[HTML]{CCD2D8}}                         & \multicolumn{1}{c|}{\cellcolor[HTML]{CCD2D8}(1)}     & \multicolumn{1}{c|}{\cellcolor[HTML]{CCD2D8}(2)} & \multicolumn{1}{c|}{\cellcolor[HTML]{CCD2D8}(3)} & (4)                                           \\ \hline
\rowcolor[HTML]{E7EAED} 
\multicolumn{1}{|l|}{\cellcolor[HTML]{E7EAED}Pre-disaster mobility} & \multicolumn{1}{c|}{\cellcolor[HTML]{E7EAED}0.0213} & \multicolumn{1}{c|}{\cellcolor[HTML]{E7EAED}0.0213}    & \multicolumn{1}{c|}{\cellcolor[HTML]{E7EAED}0.0282}    & \multicolumn{1}{c|}{\cellcolor[HTML]{E7EAED}-0.0202}                                             \\ \hline
\rowcolor[HTML]{CCD2D8} 
\multicolumn{1}{|l|}{\cellcolor[HTML]{CCD2D8}}                         & \multicolumn{1}{c|}{\cellcolor[HTML]{CCD2D8}p=0.238} & \multicolumn{1}{c|}{\cellcolor[HTML]{CCD2D8}p=0.260}    & \multicolumn{1}{c|}{\cellcolor[HTML]{CCD2D8}p=0.133}    &  \multicolumn{1}{c|}{\cellcolor[HTML]{CCD2D8}p=0.282}                                             \\ \hline
\rowcolor[HTML]{E7EAED} 
\multicolumn{1}{|l|}{\cellcolor[HTML]{E7EAED}Pre-disaster segregation}    & \multicolumn{1}{c|}{\cellcolor[HTML]{E7EAED}-}       & \multicolumn{1}{c|}{\cellcolor[HTML]{E7EAED}0.0002}    & \multicolumn{1}{c|}{\cellcolor[HTML]{E7EAED}0.0038}    & \multicolumn{1}{c|}{\cellcolor[HTML]{E7EAED}0.0273}                                              \\ \hline
\rowcolor[HTML]{CCD2D8} 
\multicolumn{1}{|l|}{\cellcolor[HTML]{CCD2D8}}                         & \multicolumn{1}{c|}{\cellcolor[HTML]{CCD2D8}-}       & \multicolumn{1}{c|}{\cellcolor[HTML]{CCD2D8}p=0.991}    & \multicolumn{1}{c|}{\cellcolor[HTML]{CCD2D8}p=0.835}    &  \multicolumn{1}{c|}{\cellcolor[HTML]{CCD2D8}p=0.100}                                             \\ \hline
\rowcolor[HTML]{E7EAED} 
\multicolumn{1}{|l|}{\cellcolor[HTML]{E7EAED}Bridge index}             & \multicolumn{1}{c|}{\cellcolor[HTML]{E7EAED}-}       & \multicolumn{1}{c|}{\cellcolor[HTML]{E7EAED}-}    & \multicolumn{1}{c|}{\cellcolor[HTML]{E7EAED}-0.0373$^{**}$}    &  \multicolumn{1}{c|}{\cellcolor[HTML]{E7EAED}-0.0378$^{**}$}                                             \\ \hline
\rowcolor[HTML]{CCD2D8} 
\multicolumn{1}{|l|}{\cellcolor[HTML]{CCD2D8}}                         & \multicolumn{1}{c|}{\cellcolor[HTML]{CCD2D8}-}       & \multicolumn{1}{c|}{\cellcolor[HTML]{CCD2D8}-}    & \multicolumn{1}{c|}{\cellcolor[HTML]{CCD2D8}p=0.041}    &  \multicolumn{1}{c|}{\cellcolor[HTML]{CCD2D8}p=0.016}                                              \\ \hline
\rowcolor[HTML]{E7EAED} 
\multicolumn{1}{|l|}{\cellcolor[HTML]{E7EAED}Centrality}               & \multicolumn{1}{c|}{\cellcolor[HTML]{E7EAED}-}       & \multicolumn{1}{c|}{\cellcolor[HTML]{E7EAED}-}    & \multicolumn{1}{c|}{\cellcolor[HTML]{E7EAED}-}    &  \multicolumn{1}{c|}{\cellcolor[HTML]{E7EAED}\textbf{0.0945$^{***}$}}                                             \\ \hline
\rowcolor[HTML]{CCD2D8} 
\multicolumn{1}{|l|}{\cellcolor[HTML]{CCD2D8}}                         & \multicolumn{1}{c|}{\cellcolor[HTML]{CCD2D8}-}       & \multicolumn{1}{c|}{\cellcolor[HTML]{CCD2D8}-}    & \multicolumn{1}{c|}{\cellcolor[HTML]{CCD2D8}-}    & \multicolumn{1}{c|}{\cellcolor[HTML]{CCD2D8}\textbf{p<0.001}}                                              \\ \hline
\rowcolor[HTML]{E7EAED} 
\multicolumn{1}{|l|}{\cellcolor[HTML]{E7EAED}Const}                    & \multicolumn{1}{c|}{\cellcolor[HTML]{E7EAED}-0.3865$^{***}$}  & \multicolumn{1}{c|}{\cellcolor[HTML]{E7EAED}-0.3865$^{***}$}    & \multicolumn{1}{c|}{\cellcolor[HTML]{E7EAED}-0.3865$^{***}$}    & \multicolumn{1}{c|}{\cellcolor[HTML]{E7EAED}-0.3865$^{***}$}                                              \\ \hline
\rowcolor[HTML]{CCD2D8} 
\multicolumn{1}{|l|}{\cellcolor[HTML]{CCD2D8}}                         & \multicolumn{1}{c|}{\cellcolor[HTML]{CCD2D8}p<0.001} & \multicolumn{1}{c|}{\cellcolor[HTML]{CCD2D8}p<0.001}    & \multicolumn{1}{c|}{\cellcolor[HTML]{CCD2D8}p<0.001}    & \multicolumn{1}{c|}{\cellcolor[HTML]{CCD2D8}p<0.001}                                              \\ \hline
\rowcolor[HTML]{E7EAED} 
\multicolumn{1}{|l|}{\cellcolor[HTML]{E7EAED}R$^2$}                       & \multicolumn{1}{l|}{\cellcolor[HTML]{E7EAED}0.0217}  & \multicolumn{1}{l|}{\cellcolor[HTML]{E7EAED}0.0217}    & \multicolumn{1}{l|}{\cellcolor[HTML]{E7EAED}0.0856}    & \multicolumn{1}{l|}{\cellcolor[HTML]{E7EAED}\textbf{0.3444}} \\ \hline
\rowcolor[HTML]{CCD2D8} 
\multicolumn{1}{|l|}{\cellcolor[HTML]{CCD2D8}Adjusted R$^2$}              & \multicolumn{1}{l|}{\cellcolor[HTML]{CCD2D8}0.0064} & \multicolumn{1}{l|}{\cellcolor[HTML]{CCD2D8}-0.0094}    & \multicolumn{1}{l|}{\cellcolor[HTML]{CCD2D8}0.0414}    & \multicolumn{1}{l|}{\cellcolor[HTML]{CCD2D8}\textbf{0.3014}} \\ \hline
\multicolumn{5}{c}{\footnotesize Note: $^{***}$ p<0.01, $^{**}$ p<0.05, $^{*}$ p<0.1.}\\
\end{tabular}%
}
\end{table}

\end{document}